\documentclass[11pt]{article}
\usepackage{cite}
\newcommand{\beq}{\begin{equation}}
\newcommand{\eeq}{\end{equation}}
\newcommand{\beqn}{\begin{eqnarray}}
\newcommand{\eeqn}{\end{eqnarray}}

\begin{document}

\begin{titlepage}
\begin{flushright}
{December 1998}
\end{flushright}
\vskip 0.5 cm

\begin{center}
            {\LARGE{\bf  Effective action for QED in 2+1 dimensions at finite temperature}} 
\vskip 0.6 cm
 Marcelo Hott\footnote{E-mail address: m.hott1@physics.oxford.ac.uk}\renewcommand{\thefootnote}{\fnsymbol{footnote}}\footnote{On leave of absence from UNESP - Campus de Guaratinguet\'a - SP - Brazil.}
, Georgios Metikas\renewcommand{\thefootnote}{\arabic{footnote}}\addtocounter{footnote}{-1}\footnote{E-mail address: g.metikas1@physics.oxford.ac.uk} \\

\vskip 0.3 cm
{Department of Physics, Theoretical Physics,\\
University of Oxford, 1 Keble Road, Oxford OX1 3NP}
\end{center}
\vskip 2 cm

\abstract  
We calculate the effective action for a constant magnetic field and a time-dependent time-component of the gauge field in 2+1 dimensions at finite temperature. We also discuss the behaviour of the charge density and the fermion condensate as order parameters of symmetry breaking. 

\end{titlepage}

\noindent{\Large \textbf{Introduction}}
\vskip 0.5cm

Many 
aspects of Quantum Electrodynamics in 2+1 dimensions (QED$_{3}$) \cite{jackiwcs}
in the presence of external fields have already been studied. The dynamical generation of a Chern-Simons (CS) term for the gauge field \cite{redlich}, the formation of a fermion
condensate in the presence of an external magnetic field \cite{miranski,koreanguys2} and the
breaking of Lorentz symmetry \cite{hosotani,koreanguys2} are some of the many features exhibited by QED$_{3}$. Some of these aspects are desirable in
3+1 dimensions and also very important in the study of 
superconductivity \cite{banks,fradkin,likken,dorey} and the quantum Hall effect \cite{wilczek} in planar systems.

In order to carry the study of such phenomena further it is important 
to consider the effects of temperature; these might sometimes be
completely different from one's expectations. Particularly, it has been 
shown that the development of the fermion condensate catalyzed by a magnetic field is unstable at finite temperature \cite{hottcond} and that the issue of gauge invariance at finite temperature is only consistently addressed, if a re-summation of the graphs is carried out \cite{dunne1,deser,schapo3}. This question has also been extensively reported using different techniques in the context of QED$_{3}$ \cite{aitchifosco,schapo4,kikuawa,gonzalez,fosco,salcedo} as well as by means of an  analogous quantum mechanical model
 which allows one to perform a comparative analysis of the perturbation series with exact results \cite{dascs,dasbarc} - in fact this was the approach used in one of the seminal works on the subject \cite{dunne1}.

Although the exact contribution of the time-component of the gauge field has been considered only for the parity violating term of the effective action, it also contributes to the parity invariant one and this may result in extra effects on quantities such as the fermion condensate and the density of charge generated dynamically. To explore such effects is the main purpose of this article. By making use of the Fock-Schwinger proper-time technique \cite{schwinger} we compute the one-loop effective action at finite temperature for a specific configuration, namely, a constant magnetic field and a time-dependent time-component of the gauge field. We also address the aspects of gauge invariance and temperature dependence of the thermal condensate and the charge density. With this programme our results go beyond those obtained in \cite{gonzalez}, where the same technique was used.

In QED the one-loop effective action for the gauge field is obtained by integrating out the fermions

\beqn
e^{i S_{eff}[A]} & = & \int {\cal D} \psi {\cal D} \bar{\psi} \; \exp{\left[ i \int d^{3}x \bar{\psi} \;  (i \not\! \partial + e \not\!{A} - m ) \psi \right] }  \nonumber \\
& \times &   \exp{ \left[ i \int d^{3}x \; ( - \frac{1}{4} F^{\mu \nu} F_{\mu \nu} + \theta \epsilon_{\mu \nu \alpha} A^{\mu} \partial^{\nu} A^{\alpha} ) \right] } \; , 
\eeqn

\noindent which can be calculated exactly only for some configurations of the field. In this case the gauge field is understood as a classical background field and many important features can be analyzed from the resulting effective action at zero \cite{schwinger,dittrich1,dunne2} or at finite temperature \cite{dittrich2,elmfors,rojas} and finite density \cite{chodos,koreanguys1}. Whenever the effective action can not be obtained exactly, it is useful to adopt some approximation. The derivative expansion technique  is such an approximation \cite{fraser} which has been successfully employed to analyze different aspects of effective actions in 1+1 \cite{daskarev} and in 2+1 dimensions \cite{aitchifoscomaz,dunne3}. 

The  CS term at finite temperature was obtained by means of the derivative expansion \cite{dasbabu} as well as other techniques \cite{niemi,ishikawa,pisarski,poppitz,luca,burgess,kim}. Although there was strong evidence that the CS coefficient for a non-abelian gauge field should be quantized at finite temperature \cite{pisarski}, it was found to depend smoothly on the temperature, leading to the conjecture that it would not depend on the temperature at all \cite{pisarski}. Later it was proved that the derivative expansion could only be used to obtain the CS term for some very special configurations of the gauge field due to the intrinsic non-locality of the CS coefficient at finite temperature, which could not be removed because of the essential non-analyticity present in the vacuum-polarization graph \cite{kaoyang,aitchifoscozuk,zuk}. More recently the subject gained new impetus in references \cite{schapo1,schapo2} where it was argued  on the basis of gauge invariance of the partition function that there was a contradiction between the quantization of the CS coefficient  at finite temperature and the results obtained at the perturbative level. 

The solution to this puzzle was given in \cite{dunne1} where a 0+1-dimensional model was used to demonstrate that the effective action at finite temperature is compatible with gauge invariance, if it is obtained exactly; as a consequence of that, it is not an extensive quantity as it was supposed to be on the grounds of the perturbative expansion. Since it is impossible to sum all the graphs for the general problem in 2+1 dimensions, one has to choose some specific configurations of the background gauge field. It was shown in references  \cite{deser,schapo3} that a configuration with the space-components of the gauge field depending only on the space-coordinates and the time-component depending on the Euclidean time, namely

\beq
A_{j}  =   A_j(\vec{x}), \; \; \; \; A_{3}   =   A_{3}(\tau), \label{config}
\eeq 

\noindent allows one to obtain the compatibility between gauge invariance and the CS term generated dynamically.

There it was also emphasized that in Euclidean time there is room to define homotopically non-trivial gauge transformations, as for instance the one given by

\beq
\Lambda (\tau ) = - \int_{0}^{\tau} d\tau^{\prime} A_{3}(\tau^{\prime} ) + \left( \frac{1}{\beta} \int_{0}^{\beta} d\tau^{\prime} A_{3}(\tau^{\prime} ) + \frac{2 \pi k}{e \beta} \right) \tau \; ,
\eeq
 
\noindent where $k$ is an integer number which defines the homotopically distinct configurations. Under this transformation one notes that the time-component of the gauge field becomes constant

\beqn
A_{3}(\tau ) \rightarrow   {\tilde A}_{3}(\tau ) & = & A_{3}(\tau ) + \partial_{\tau } \Lambda (\tau)  \nonumber \\
& = & \frac{1}{\beta} \int_{0}^{\beta} d\tau^{\prime} A_{3}(\tau^{\prime} ) + \frac{2 \pi k}{e \beta} \; ,  \label{A3}
\eeqn

\noindent and the space-components remain invariant. One can also check that this gauge transformation on the fermion fields

\beq
\psi (\vec{x} , \tau ) \rightarrow {\tilde \psi} (\vec{x} , \tau ) = e^{i e \Lambda (\tau ) } \psi (\vec{x} , \tau )
\eeq

\noindent is also compatible with the anti-periodic time boundary condition

\beq
\psi (\vec{x} , 0) = - \psi (\vec{x} , \beta ).
\eeq

\noindent In fact

\beq
\Lambda (\beta ) = \Lambda (0) + \frac{2 \pi k}{e}
\eeq
\bigskip

\noindent{\Large \textbf{ Charge density and effective action }}
\vskip 0.5cm

We are going to adopt the same gauge configuration as given by equation (\ref{config}) but for a constant magnetic field $F_{12}=B$

\beq
A_{j}= \frac{1}{2} F_{jk} x^{k},
\eeq

\noindent for which it is possible to obtain an exact result for the fermion propagator. This can be obtained from the covariant calculation \cite{redlich} and made particular for our purpose. It is given in Minkowski space-time by

\beqn
G(x,x^{\prime}) & = & i \frac{e^{3 \pi i 4}}{8 \pi^{3/2}} \Phi (x,x^{\prime}) \int_{0}^{\infty} \frac{ds}{s^{3/2} } \; e^{- i m^{2} s}  \nonumber \\
& \times &  \exp { \left( i \left[ e A_{0} (t - t^{\prime}) - \frac{(t - t^{\prime})^2 }{4 s} + \frac{e |B| s}{4 s} \cot{(e |B| s)} ({\rm \vec{x} - \vec{x}^{\prime }})^{2} \right] \right) }  \nonumber \\
& \times &  \left\{ m - \gamma^{0} \; \frac{(t - t^{\prime})}{2 s} + \frac{e |B| s}{2 s} \; \epsilon_{jk} \gamma^{j} {\rm (x - x^{\prime})}^{k} + \frac{e |B| s}{2 s} \cot{(e |B| s)} \; \vec{\gamma}. \vec{\rm{x}} \right\}  \nonumber \\
& \times &  \frac{e |B| s}{\sin{(e |B| s)}} \left[ \cos{(e |B| s)} - i \gamma^{0} \frac{\ ^*F_0}{|B|} \sin{(e |B| s)} \right]  \; , \label{coordprop}
\eeqn

\noindent where $\Phi (x,x^{\prime})$ is the gauge dependent factor

\beq
\Phi (x,x^{\prime}) = \exp{\left(i e  \int_{x^{\prime} }^{x} d \eta^{\mu} A_{\mu}(\eta) \right) } \; 
\eeq

\noindent and ${\ ^*F_0} = (1/2) \epsilon_{jk} F_{jk} $. One can note that a simple gauge transformation on this factor removes the $A_{0}=constant$ taken into account explicitly in the propagator. However we are going to see that at finite temperature there are factors dependent on the time-component of the gauge field that can not be removed by a gauge transformation. 

To implement the finite temperature calculation in the imaginary-time formalism we recall that the propagator can also be rewritten as

\beq    
G(x,x^{\prime}) =  \Phi (x,x^{\prime}) \int \frac{d^{3}k_{E}}{(2 \pi )^{3}} e^{-ik_{E}(x - x^{\prime})} \tilde{G}(k_{E}) \; ,
\eeq

\noindent where

\beqn
\tilde{G}(k_{E}) & = & - \int_{0}^{\infty} ds \; \exp{\left(- i s \left[ m^{2} + (k_{3} + e {\tilde A}_{3})^{2} + {\vec{k}}^{2} \frac{\tan{ (e |B| s)} }{e |B| s} \right] \right) }  \nonumber \\
& \times &  \left\{ m - \gamma_{3} (k_{3} - e {\tilde A}_{3} ) - \vec{\gamma}.\vec{k} + \epsilon_{ij} \gamma^{i} k^{j} \tan{ (e |B| s) } \right\}  \nonumber \\
& \times &  \left[1 + \gamma_{3} \frac{\ ^*F_3}{|B|} \tan{(e |B| s)} \right]  \; ,
\eeqn

\noindent is its Fourier-transform in Euclidean version ($\gamma_{3} = - i \gamma_{0}$) and ${\tilde A}_3$ is given by (\ref{A3}). Once again we note that a simple redefinition of $k_{3}$ as a continuous variable renders $\tilde{G}(k_{E})$ gauge-independent, leaving the gauge dependence only on the factor  $\Phi (x,x^{\prime})$.
\medskip

From now on we consider only the gauge invariant propagator in order to compute gauge invariant quantities, for example the charge density which is given by

\beq
<j_{3}> =  ie \; tr[\gamma_{3} G_{E}(x,x^{\prime})] \vert_{x=x^{\prime}} . \eeq

Performing the trace of the $\gamma$-matrices and the integration over $\vec{k}$, we obtain the following expression for the charge density,

\beqn
<j_{3}> & = & \frac{e^{2} }{2 \pi} \frac{1}{\beta } \sum_{n=-\infty}^{\infty} \int_{0}^{\infty} ds \exp{\left(-is [ m^{2} + (w_{n} + e {\tilde A}_{3})^{2}] \right) }  \nonumber \\
& \times &  [m {\ ^*F_3}  - |B|(w_{n} + e {\tilde{A}}_{3}) \coth{(e |B| s)}] \; ,  
\eeqn

\noindent where $w_{n}= {(2n + 1) \pi / \beta }$. We can see from the above expression that there are two contributions to the expectation value of the charge density, the parity violating (PV) one which is proportional to $ \ ^*F_3 $ and the parity invariant (PI) one which is proportional to  $ |B| $. Now we can obtain the parity violating charge density, namely

\beqn
<j_{3}>^{PV} & = & - i m \frac{e^{2} {\ ^*F_3}}{2 \pi} \frac{1}{\beta} \sum_{n=-\infty}^{\infty} \frac{1}{ [m^{2} + (w_{n} + e {\tilde A}_{3})^{2}]}  \nonumber \\
& = &  -i \frac{m}{|m|} \frac{e^{2} {\ ^*F_3}}{4 \pi} \tanh{\left( \frac{|m| \beta}{2} \right) } \left[ \frac{1}{1 + \tanh^{2}{\left( \frac{|m| \beta }{2} \right) } \tan^{2}{\left( \frac{e {\tilde A}_{3} \beta }{2} \right) } } \right]  \nonumber \\
& \times &  \left[ 1 + \tan^{2}{\left( \frac{e {\tilde{A}}_3 \beta}{2} \right) } \right] \; . \label{j3pv1} 
\eeqn

\noindent We note that this expression is indeed gauge invariant under both small and large gauge transformations.

Since this term comes from the existence of a fermion mass it is important to ask what happens to the charge density generated dynamically in the limit $m \rightarrow 0$. From  the above expression we find that it vanishes independently of ${\tilde A}_3$ for any temperature, although it survives at zero temperature \cite{redlich}. This phenomenon also happens to the fermion condensate in the reducible representation \cite{hottcond} and will also be seen to happen here with the irreducible representation. The melting of the charge condensation at finite temperature can clearly be seen by rewriting expression (\ref{j3pv1}) as

\beq
<j_{3}>^{PV} =  - i \frac{m}{|m|} \frac{e^{2} {\ ^*F_3}}{4 \pi} \left[ 1 - \frac{1}{e^{\beta (|m| + i e {\tilde A}_{3})} + 1} -  \frac{1}{e^{\beta (|m| - i e {\tilde A}_{3})} +1 } \right] \; . \label{j3pv2}
\eeq

\noindent We can see that the $T=0$ contribution survives in the limit  $m \rightarrow 0$ but it is canceled by the thermal fluctuations. Equation (\ref{j3pv2}) also highlights the r\^ole played by $i e {\tilde{A}}_{3}$ as a chemical potential.

From equation (\ref{j3pv1}) we can obtain the parity violating contribution to the effective action by recalling that

\beq
<j_{3}> = - \frac{\delta S_{eff}}{\delta A_{3}(\tau )} \; . \label{j3A3}
\eeq
\newpage
\noindent Then we obtain

\beq
 S_{eff}^{PV} = \frac{i e }{2 \pi } \frac{m}{|m|} \int d^2x \ {\ ^*F_3}  \arctan{\left[ \tanh{\left( \frac{|m| \beta}{2} \right) } \tan{\left( \frac{e {\tilde A}_{3} \beta}{2} \right) } \right]} \; . \label{spv}
\eeq

\noindent This is the result found in references \cite{deser,schapo3}. It can also be rewritten in a more suitable form to check the consequences of a gauge transformation, 

\beqn
S_{eff}^{PV}&=& \frac{e }{4 \pi } \frac{m}{|m|} \int d^2x \ {\ ^*F_3}
 \left\{ i e \int_{0}^{\beta} d\tau \ A_3(\tau) + 2 \pi i k \right. \nonumber
\\ & - &  \left. \ln{ \left[ 1 + e^{-\beta (|m|  - i e {\tilde A}_{3}
)} \right] } +  \ln{\left[ 1 + e^{- \beta(|m|  + i e {\tilde A}_{3})} \right] } \right\} \ . 
\eeqn

\noindent We point out that the superscript PV has been used for the
effective action only for making clear that it is associated with the
parity violating charge density. It is the effective Lagrangian which
possesses the parity violating property and not the effective action. The latter is invariant under small gauge transformation ($k=0$) and changes by ${i e k \Phi / 2}$ under large gauge transformation, where $\Phi$ is the magnetic flux. The change under large gauge transformation comes from the zero-temperature contribution and, if the magnetic flux is quantized in units of ${4 \pi / e}$, the partition function for the gauge field is invariant under large gauge transformation. Furthermore one can also note that this is not the CS term at finite temperature; this is hidden in the factors with logarithmic functions which are not extensive quantities. 

The parity invariant contribution to the charge density is given by

\beqn
<j_{3}>^{PI} & = & - \frac{e^{2} |B| }{2 \pi} \frac{1}{\beta } \sum_{n=-\infty}^{\infty} \int_{0}^{\infty} ds \; e^{-i s [m^{2} + (w_{n} + e {\tilde A}_{3})^{2}]} (w_{n} + e {\tilde A}_{3}) \cot{(e |B| s)}  \nonumber \\
& = & -  i \; \frac{e^{2} |B| }{4 \pi} \sum_{n=0}^{\infty} \sum_{s=1}^{2} \left[ \frac{1}{e^{\beta(E_{n,s} + i e {\tilde A}_{3})} + 1} - \frac{1}{e^{\beta(E_{n,s} - i e {\tilde A}_{3})} + 1} \right] \; ,  \label{j3pi}
\eeqn

\noindent where

\[ E_{n,s} = \sqrt{m^{2} + 2e|B|(n + s - 1)} \; , \]

\noindent is the energy of the Landau levels.

We note that, at $T=0$, the parity invariant charge density vanishes, as expected \cite{redlich, schwinger}. More interestingly, it vanishes in the limit $ {\tilde A}_{3} = 0 $, as well. Comparing equations (\ref{j3pv2}) and (\ref{j3pi}) we see that the parity violating contribution to the charge density is proportional to $ ^*F_3$ and only the lowest Landau level contributes to it; on the other hand the parity invariant part is proportional to $|B|$ and all the Landau levels contribute. It is also important to remember that the parity violating contribution is a peculiar feature of QED in an odd number of space-time dimensions with fermions in the irreducible representation; on the other hand the parity invariant one exists in both the irreducible and the reducible representation of QED$_3$. Actually, it is the only contribution in the reducible representation, as is expected, since QED$_3$ in the reducible representation is very similar to QED$_4 $. 
\bigskip
\newpage

Using formulae (\ref{j3A3}) and (\ref{j3pi}) we can obtain the parity invariant effective Lagrangian, which is given by

\beqn
 {{\cal L}_{eff}}^{PI} & = & \frac{e |B|}{4 \pi } \sum_{n=0}^{\infty} \sum_{s=1}^{2} \left\{  E_{n,s} + \frac{1}{\beta} \ln{\left[ 1 + e^{-\beta (E_{n,s} + i e {\tilde A}_{3})} \right] } \nonumber \right. \\
& + & \left.  \frac{1}{\beta} \ln{\left[ 1 + e^{- \beta (E_{n,s} - i e {\tilde A}_{3})} \right] } \right\}  \; . \label{lefft} 
\eeqn

\noindent The first term in the curly brackets is the zero temperature contribution and can be rewritten as

\beq
 {{\cal L}_{eff}}^{PI}(T=0) = \frac{e^{i3 \pi / 4}}{8 \pi^{3/2}} \int_{0}^{\infty} \frac{ds}{s^{5/2}} e^{-i m^{2} s} e |B| s \cot{(e |B| s)} \; .
\eeq

\noindent Finally, we note that this effective Lagrangian has to be properly renormalized. The result can be read off directly from references \cite{dittrich1,elmfors}
\beq
{{\cal L}_{eff}}^{PI}(T=0) =  \frac{e^{i3 \pi / 4}}{8 \pi^{3/2}} \int_{0}^{\infty} \frac{ds}{s^{5/2}} e^{-i m^{2} s} \left(e |B| s \cot{(e |B| s)} -1 - \frac{1}{3} (e s |B|)^{2} \right) \; ,
\eeq

\noindent and the whole effective Lagrangian is given by

\beq
{\cal L}_{eff} = - \frac{1}{2} {\ ^*F_3}^2 - \theta {\ ^*F_3}A_{3} + {{\cal L}_{eff}}^{PV} +{{\cal L}_{eff}}^{PI}. 
\eeq
\noindent{\Large \textbf{Fermion condensate}}
\vskip 0.5cm

Another quantity which is significant in the analysis of symmetry breaking is the fermion condensate. It can be shown to be given by

\beq
<\bar{\psi} \psi> =   i \; tr[G_{E}(x,x^{\prime})] \vert_{x=x{\prime}} \; .
\eeq
    
\noindent It also has two contributions which will be evaluated by means of the well-known formula 
\beq
<\bar{\psi} \psi>= - \frac{\partial {\cal L}_{eff}}{\partial m} \; . \label{dldm}
\eeq 
 \noindent  The first one, which comes from the parity invariant effective Lagrangian, is also present in QED$_3$ with fermions in the reducible representation and is given by  

\beqn
<\bar{\psi} \psi>^{PI} & = & - m 
\frac{e |B|}{2 \pi} \frac{1}{\beta } \sum_{n=-\infty}^{\infty} \int_{0}^{\infty} ds \exp{\left(-i s [m^{2} + (w_{n} + e {\tilde A}_{3})^{2}] \right)}  \cot{(e |B| s)}   \nonumber \\
& = &  - m  \frac{e |B|}{4 \pi} \sum_{n=0}^{\infty} \sum_{s=1}^{2} \frac{1}{E_{n,s}} \left[ 1 - \frac{1}{e^{\beta(E_{n,s} + i e {\tilde A}_{3})} + 1} - \frac{1}{e^{ \beta(E_{n,s} - i e {\tilde A}_{3})} + 1} \right] \; . \nonumber \\
\label{cond1} 
\eeqn

\noindent This expression is the same found in \cite{hottcond} using
real-time formalism, apart from a factor of $1/2$ due to the trace of
gamma matrices. Therefore the analysis carried out there can naturally
be applied here. First we note that all Landau levels contribute to
this part of the condensate and, although it is non-vanishing in the
limit $m \rightarrow 0$ at zero temperature \cite{miranski}, it melts
at any finite temperature for $m \rightarrow 0$ independently of
$A_{3}$. Again we note that equation (\ref{cond1}) shows a similarity
between  the $i e {\tilde A}_3$ and a chemical potential with the
temperature taken into account. In the reducible representation this
is the only contribution to the order parameter. In \cite{farakos} a
$B$-dependent critical temperature was found for the case of small but
non-vanishing mass. Above this critical temperature and in the regime
$m^{2} \ll T^{2} \ll eB$ the condensate vanishes. 

The second contribution to the condensate comes from the parity violating effective Lagrangian and reads

\beq
<\bar{\psi} \psi>^{PV} = - \frac{e {\ ^*F_3}}{4 \pi} \frac{1}{\beta } \sum_{n=-\infty}^{\infty} \int_{0}^{\infty} ds \exp{ \left(-i s [m^{2} + (w_{n} + e {\tilde A}_{3})^{2}] \right) }  (w_{n} + e {\tilde A}_{3}) \; .
\eeq

\noindent It is easy to see that it can be recast as follows

\beqn
<\bar{\psi} \psi>^{PV} & = &  - i  \frac{ e { \ ^*F_3 } }{4 \pi} \tan{ \left( \frac{e {\tilde A}_{3} \beta}{2} \right) } \left[ \frac{1}{1 + \tanh^{2}{ \left( \frac{|m| \beta}{2} \right) } \tan^{2}{(\frac{e {\tilde A}_{3} \beta}{2})}} \right]  \nonumber \\
& \times &  \left[ 1 - \tanh^{2}{ \left( \frac{|m| \beta}{2} \right) } \right] \nonumber \\ 
 & = & - \frac{ e { \ ^*F_3 } }{4 \pi} \left\{ \frac{1}{ e^{ \beta ( |m| - i e \tilde{A}_{3} ) } + 1 } - \frac{1}{ e^{ \beta ( |m| + i e \tilde{A}_{3} ) } + 1 } \right\} .  
\eeqn

\noindent We note that it vanishes, if $ T=0 $ or $ \tilde{A}_{3} = 0 $. Finally, it can be shown that only the lowest Landau level contributes to this part of the condensate. 
\bigskip

\noindent{\Large \textbf{Conclusions}}
\vskip 0.5cm

We have found the thermal effective action for a particular
configuration of the gauge field, namely $ A_{j}= \frac{1}{2} F_{jk}
x^{k}$ with $F_{12}=B=$constant and $A_{3}=A_{3}(\tau )$. Due to the
presence of $A_{3}(\tau )$ we found that the associated partition function is gauge invariant. In the course of our derivation we also calculated the charge density and the fermion condensate, highlighting their origins in terms of the contribution of the Landau levels and the time-component of the gauge field. Although we found these quantities very unstable in the zero-limit of the relevant parameters, the fermion mass and $A_{3}(\tau )$ analysis developed here can be useful whenever quantum fluctuations of the gauge field are taken into account. We also pointed out the similarities of our results with those obtained at finite density. More investigation on these similarities, their physical meaning and consequences are under study and will be reported elsewhere.
\bigskip 

\noindent{\Large \textbf{Acknowledgements}}
\vskip 0.5cm

M. Hott is supported by Funda\c c\~ao de Amparo a Pesquisa do Estado de
S\~ao Paulo (FAPESP-Brazil). G. Metikas is grateful to PPARC (UK) for
financial support. The authors wish to thank Prof. I.J.R Aitchison,
Dr. N.E. Mavromatos and Dr. A. Momen for valuable discussions.   
\newpage


\end{document}